\documentclass[
aps,prb,superscriptaddress,twocolumn,floatfix,10pt,
longbibliography
]{revtex4-2}
\usepackage[utf8]{inputenc}
\usepackage{amsmath,amssymb}
\usepackage{empheq}
\usepackage{float}
\usepackage{lipsum}
\usepackage{mathtools,cuted}
\usepackage{esvect}
\usepackage{multirow}
\usepackage{xcolor}
\usepackage{soul}
\usepackage[export]{adjustbox}
\usepackage{graphicx}
\usepackage{dcolumn}
\usepackage{bm}
\usepackage{hyperref}
\usepackage{physics}
\usepackage[mathlines]{lineno}
\usepackage[capitalise]{cleveref}
\usepackage{todonotes}
\usepackage{bbm}
\usepackage{orcidlink}

\newcommand{\kk}[0]{\mathbf{k}}

\begin{document}
	
	\title{Universality of dimensional crossovers in topological insulators}
	
\author{L. Eek}
\affiliation{Institute of Theoretical Physics, Utrecht University, Utrecht, 3584 CC, Netherlands}
\author{Z.~F.~Osseweijer}
\affiliation{Institute of Theoretical Physics, Utrecht University, Utrecht, 3584 CC, Netherlands}
\author{C. Morais Smith}
\affiliation{Institute of Theoretical Physics, Utrecht University, Utrecht, 3584 CC, Netherlands}
 
	\date{\today}
	
	\begin{abstract}
    We investigate dimensional crossovers in minimal tight-binding models of three-dimensional (3D) topological insulators subject to geometric confinement. While thin films are commonly understood to host a crossover from a 3D strong topological insulator to a two-dimensional (2D) quantum spin Hall phase via hybridization of surface states, we demonstrate that this picture is incomplete once bulk confinement effects and boundary termination are fully taken into account. Using lattice models, we show that reducing the system size induces a strongly non-monotonic dependence of the topology on thickness and microscopic parameters, leading to a sequence of topological phase transitions that is highly sensitive to surface termination. In particular, we find a cascade of dimensional reduction from a 3D topological insulator to a 2D quantum spin Hall phase and ultimately to a one-dimensional phase consisting of end states of Kramers pairs protected by inversion symmetry. Remarkably, we show that both the 2D and 1D topological phases can emerge even when the corresponding 3D bulk phase is topologically trivial. Our results reveal an unexpected universality in the phase diagrams of 3D-to-2D and 2D-to-1D crossovers, pointing toward a unified framework for topology under dimensional reduction.
	\end{abstract}
	
	\maketitle{}
	

\textit{Introduction---}
Topological insulators are classified according to their symmetries and dimensionality, as captured by the tenfold way \cite{altland1997nonstandard, schnyder2008classification, ryu2010topological}. While this classification is formulated for bulk systems, reducing the system size can qualitatively modify their topological properties. In particular, thin films of 3D topological insulators provide a natural setting in which a transition from a three-dimensional strong topological insulator (3DTI) to a 2D quantum spin Hall (QSH) phase can occur \cite{zhang2010crossover, xu2015unconventional, wang2019dimensional, liu2010oscillatory, van2025observation, moes2024characterization}.\\
The common interpretation of this crossover relies on hybridization between Dirac surface states on opposite surfaces of the film. Within this picture, the overlap of the surface wave functions opens a gap at the Dirac point, which may drive the system into a QSH phase. At a phenomenological level, this mechanism can be captured by introducing a tunneling term between the two surfaces in an effective Bernevig-Hughes-Zhang (BHZ) model \cite{bernevig2006quantum, shan2010effective, asmar2018topological, maisel2024topology}. This term gaps the surface spectrum and can, depending on its sign, yield either a trivial or a topological phase. While this description captures key features of the dimensional crossover, it treats the surface states mostly independently of the bulk. In sufficiently thin films, however, quantum confinement also strongly modifies the bulk spectrum, which in turn influences the nature of the low-energy states. As a result, a complete description of the crossover requires going beyond a purely surface-based picture.\\
Dimensional crossovers can also be realized when 2D systems are confined into quasi-one-dimensional geometries. Examples include coupled Majorana chains, finite-width ribbons of Chern insulators on square and honeycomb lattices, and QSH insulators \cite{wakatsuki2014majorana,cook2023finite,traverso2024emerging,klaassen2025realization,eek2025electric,osseweijer2026topology}. While transitions from three to two dimensions are relatively straightforward within the tenfold way classification, especially when the system remains topological in both dimensions, as e.g. in class AII \cite{altland1997nonstandard}, the crossover to one dimension is more subtle. In particular, neither class A (relevant for Chern insulators) nor class AII (3DTI and QSH) supports nontrivial topology in one dimension. As a result, realizing a nontrivial 1D limit requires the emergence of additional symmetries, such as an effective chiral symmetry or, beyond the tenfold way, crystalline symmetries that protect the lower-dimensional topological phase. While there exist works on dimensional crossovers from a tight-binding perspective \cite{pertsova2014probing, mao2011tight, flores2023time}, these are typically model-specific or rely on simplified assumptions that do not fully capture the interplay between geometry, boundary conditions, and bulk band structure. 
In this work, we show that even in minimal 3DTI models, dimensional crossovers naturally emerge, and that the relation between geometrical parameters such as thickness and microscopic model parameters is highly nontrivial and strongly non-monotonic. Moreover, we demonstrate that the resulting phase structure depends sensitively on the choice of surface termination. We find that a cascade of transitions can occur upon reducing the dimensionality, from a 3DTI to a 2D QSH phase and ultimately to a 1D phase consisting of end states of Kramers pairs, where the protecting symmetry is provided by inversion symmetry. We show that the 1D and 2D lower-dimensional topological phases can arise even in regions of parameter space where the original 3D bulk model is topologically trivial. Finally, we observe a degree of universality in the resulting phase diagrams for both the 3D to 2D and 2D to 1D crossovers, raising the question of whether a unified description of dimensional reduction in topological band structures can be formulated.
\begin{figure*}
    \centering
    \includegraphics[width=\linewidth]{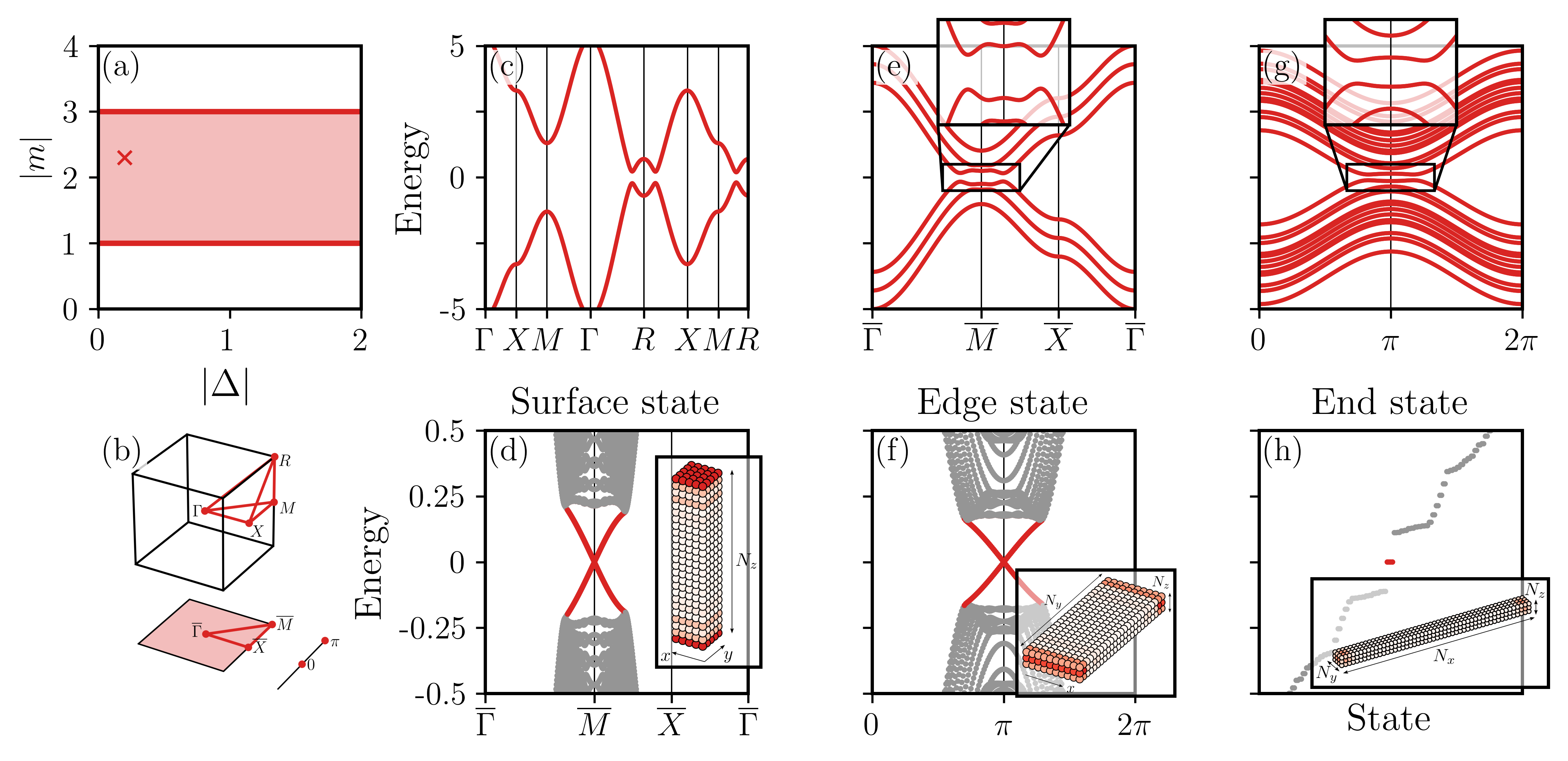}
    \caption{
(a) Phase diagram of the model as a function of $m$ and $\Delta$, with the cross indicating the parameters used throughout the rest of the figure ($t=1$). Red (white) indicates a topological phase with $\nu_0=1$ (trivial phase with $\nu_0=0$).
(b) High-symmetry path in the bulk Brillouin zone, the (100) surface Brillouin zone, and the corresponding 1D projected Brillouin zone along [10].
(c) Bulk band structure of Eq.~\eqref{eq:model} in the topological phase, exhibiting a characteristic camelback feature near the $R$ point.
(d) Slab spectrum for $[100]$ termination with thickness $N_z=25$, showing a surface Dirac cone at $\overline{M}$ (inset: surface states).
(e) Slab spectrum for $N_z=3$, where finite-size effects gap out the surface Dirac cone at $\overline{M}$.
(f) Rod spectrum with open boundary conditions in $y$ and $z$ ($N_y=25$, $N_z=3$) and periodic boundary conditions along $x$, revealing counterpropagating edge states within the bulk gap, characteristic of a 2d QSH phase (inset: QSH edge states).
(g) Rod spectrum for $N_y=4$, where the QSH edge modes are gapped out, yielding a small inverted gap at $k=\pi$.
(h) Fully open geometry with large $N_x$, ($N_y=4$, $N_z=3$), showing four in-gap modes: a spin-up/spin-down pair localized at each end of the rod.
}
    \label{fig1}
\end{figure*}



\textit{Model---}
We consider the Bloch Hamiltonian of a class-AII 3DTI on a cubic lattice,
\begin{align} \label{eq:model}
    \mathcal{H}(\kk) &=
    \bigg[m+t\sum_{i\in\{x,y,z\}} \cos (k_i) \bigg] \tau_z \sigma_x  \\
    &+ \Delta \big[\sin (k_x) \tau_z \sigma_y + \sin (k_y) \tau_z \sigma_z + \sin (k_z) \tau_x \sigma_0\big].\notag
\end{align}
In the remainder of this letter, we set the hopping $t=1$, such that every energy is measured in units of $t$. Moreover, $m$ is a mass parameter, which allows us to tune the topological phase, and $\Delta$ is a spin-orbit like hopping.
Equation~\eqref{eq:model} has time-reversal symmetry $\mathcal{T} = \mathcal{U} \mathcal{K}$ with $\mathcal{U} = \tau_y\sigma_z$ and $\mathcal{K}$ denoting complex conjugation, such that $\mathcal{H}(\kk) = \mathcal{U} \mathcal{H}^*(-\kk) \mathcal{U}^\dagger$. Furthermore, it has an inversion symmetry $\mathcal{I}=-\tau_z \sigma_x$ such that $\mathcal{H}(\kk) =\mathcal{I} \mathcal{H}(-\kk) \mathcal{I}$.\\
At the time-reversal-invariant momenta (TRIM) $\boldsymbol{\Lambda}_i=(n_x\pi,n_y\pi,n_z\pi)$ with $n_i\in\{0,1\}$, the occupied Bloch states $|u_{n} (\kk)\rangle$ can be chosen as eigenstates of the inversion operator, $\mathcal{I}|u_{n}(\boldsymbol{\Lambda}_i)\rangle=\lambda_{n}(\boldsymbol{\Lambda}_i)|u_{n}(\boldsymbol{\Lambda}_i)\rangle$, where $\lambda_{n}(\boldsymbol{\Lambda}_i)=\pm1$. Because $[\mathcal{I},\mathcal{T}]=0$, the two states of a Kramers pair share the same inversion eigenvalue. The $\mathbb{Z}_2$ invariant can therefore be evaluated from a single state in each Kramers pair, yielding the Fu-Kane formula for the strong topological invariant
\begin{equation}\label{eq:inv3D}
(-1)^{\nu_0}
=
\prod_{\boldsymbol{\Lambda}_i}
\prod_{n}^{N_{\mathrm{occ}}/2}
\lambda_{2n}(\boldsymbol{\Lambda}_i),
\end{equation}
where $N_\mathrm{occ}$ is the number of occupied bands. For the topological insulator under consideration, the system enters a 3D nontrivial phase ($\nu_0=1$) for $1<|m/t|<3$ and $|\Delta|>0$, as shown in Fig.~\ref{fig1}(a). The corresponding bulk spectrum of Eq.~\eqref{eq:model}, evaluated at parameters indicated by the cross, is displayed in Fig.~\ref{fig1}(c). The high-symmetry path through the Brillouin zone is depicted in Fig.~\ref{fig1}(b). A characteristic camelback feature is observed near the $R$ point of the bulk Brillouin zone. Upon considering a slab geometry with $[100]$ termination and thickness $N_z=25$, see Fig.~\ref{fig1}(d), a surface Dirac cone emerges at the $\overline{M}$ point of the surface Brillouin zone, giving rise to the surface states shown in the inset.
\begin{table}[]
\begin{center}
\begin{tabular*}{\columnwidth}{@{\extracolsep{\fill}} c c c c}
\hline
Class & 1D & 2D & 3D \\
\hline\hline
AII \phantom{+ $\mathcal{I}$}
& $0$ 
& $\mathbb{Z}_2$ 
& $\mathbb{Z}_2$ \\

AII + $\mathcal{I}$ 
& $\mathbb{Z}_2$ 
& $\mathbb{Z}_2$ 
& $\mathbb{Z}$ \\
\hline
\end{tabular*}
\end{center}
\caption{Topology in class AII with and without inversion symmetry ($\mathcal{I}$) across dimensions.}\label{tab:AII}
\end{table}
\begin{figure*}
    \centering
    \includegraphics[width=\linewidth]{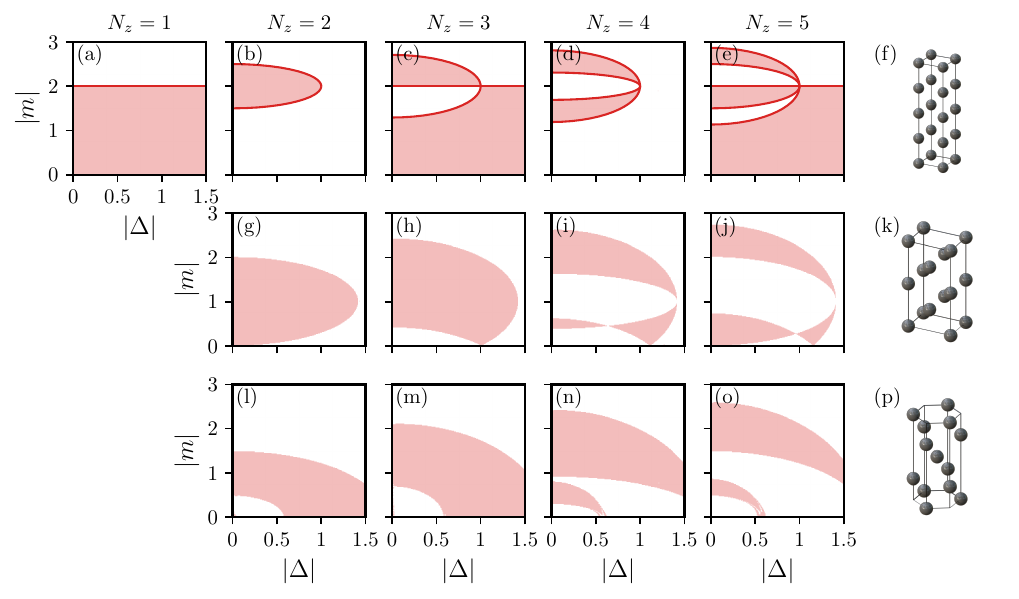}
    \caption{
QSH phase diagrams $\nu_\mathrm{QSH}$ as a function of $m$ and $\Delta$ for different surface terminations and slab thicknesses. 
(a)–(e) $[100]$ termination for $N_z=1$–$5$.
(f) Corresponding slab unit cell for $N_z=5$.
(g)–(j) $[110]$ termination for $N_z=2$–$5$.
(k) Corresponding slab unit cell for $N_z=5$.
(l)–(o) $[111]$ termination for $N_z=2$–$5$.
(p) Corresponding slab unit cell for $N_z=5$.
Red (white) regions indicate the topological (trivial) phase with $\nu_\mathrm{QSH}=1$ ($0$). In (a)-(e), analytical phase boundaries are indicated with red lines. The expressions for these boundaries are given in the End Matter.
}\label{fig2}
\end{figure*}
Considering the slab Hamiltonian introduces an additional parameter, $N_z$. For large $N_z$, the spectrum approaches the bulk limit, whereas for smaller $N_z$ confinement and hybridization effects become significant. In Fig.~\ref{fig1}(e), we show the spectrum for the same parameters as in Fig.~\ref{fig1}(d), but with $N_z=3$. In contrast to the large-$N_z$ case, the surface states have now become gapped, exhibiting a camelback feature at the $\overline{M}$ point. Solving the Hamiltonian in a rod geometry, i.e., with open boundary conditions in the $y$ and $z$ directions (with $N_y$ large) and periodic boundary conditions along $x$, reveals counterpropagating edge states within the two-dimensional bulk gap, as shown in Fig.~\ref{fig1}(f). In fact, the system has transitioned into a 2d QSH phase.\\
Repeating this procedure, we compute the rod band structure for $N_y=4$, shown in Fig.~\ref{fig1}(g) for the same parameters and $N_z$ as in Fig.~\ref{fig1}(f). The gap-traversing QSH edge modes are now gapped out, and a small inverted gap appears around $E=0$ (see inset). Solving the fully open system, with open boundary conditions in all directions and large $N_x$, reveals four in-gap modes: a spin-up/spin-down pair localized at each end of the rod, see Fig.~\ref{fig1}(h).\\
While the emergence of a QSH phase in the 2D limit is not unexpected, given that both systems belong to symmetry class AII in the Altland–Zirnbauer classification \cite{altland1997nonstandard}, the appearance of zero-energy end states in the 1D limit is more surprising. In 1D, class AII alone does not support a nontrivial topological phase. This apparent contradiction is resolved by recalling that the system also preserves inversion symmetry. The combination of time-reversal and inversion symmetry enriches the classification, leading to a phase that hosts a Kramers pair of 0D boundary modes at each end of the system. This phase is characterized by a $\mathbb{Z}_2$ invariant in one dimension (see Table~\ref{tab:AII}), which protects the presence of these end states. In the following, we first analyze the properties of the QSH phase in detail, before turning to the inversion-symmetry-protected 1D limit.

\textit{Transition to the QSH phase---} 
We now focus on the transition from the 3DTI phase to the 2D QSH phase. A variety of terminations are possible for the simple cubic lattice; in Fig.~\ref{fig1}, we considered only the $[100]$ termination. Here, we provide a detailed analysis of all low-index surfaces: $[100]$, $[110]$, and $[111]$. Certain terminations may break inversion symmetry, rendering inversion-based topological invariants, such as the Fu-Kane formula, inapplicable. Instead, we employ the Wilson-loop method of Ref.~\cite{WilsonZ2} to compute the QSH $\mathbb{Z}_2$ invariant, $\nu_\mathrm{QSH}$, from the slab Hamiltonian $\mathcal{H}_\mathrm{slab}(k_1,k_2)$ (see End Matter for additional details).\\
The QSH phase in thin 3D TI slabs arises from the hybridization of the top and bottom surface Dirac cones. Since the resulting gap scales with the overlap of these states, and hence with slab thickness, we focus on the thin-slab regime. We begin with the $[100]$ termination, with results shown in Figs.~\ref{fig2}(a)–(e); the corresponding slab unit cell for $N_z=5$ is displayed in Fig.~\ref{fig2}(f).\\
In the monolayer limit, Fig.~\ref{fig2}(a) ($N_z=1$), the system is in the QSH phase for $|m|<2$, as expected, since this limit effectively realizes the 2D BHZ \cite{bernevig2006quantum} (or spinful QWZ \cite{qi2006topological}) model on a lattice. Notably, a topological region also persists for $N_z=2$, see Fig.~\ref{fig2}(b). While stacking two QSH layers might naively suggest a trivial phase due to the $\mathbb{Z}_2$ nature of the QSH effect, interlayer coupling qualitatively modifies the low-energy physics. As shown in the End Matter, the resulting effective model remains in the QSH phase, explaining the survival of topology at $N_z=2$. Upon increasing the slab thickness [Figs.~\ref{fig2}(c)–(e)], the QSH invariant $\nu_\mathrm{QSH}$ exhibits an alternating pattern of nested ellipses centered at $(\Delta,m)=(0,2)$, with additional structure emerging as $N_z$ increases. Furthermore, for odd $N_z$ a large topological region persists for $|m|<2$, whereas for even $N_z$ this region is absent. Notably, the phase boundaries of the different phases for the [100] termination can be analytically derived (see End Matter). The analytical curves are drawn in red in Figs.~\ref{fig2}(a)–(e). Saliently, the phase diagrams for [100]-terminated 3DTIs reducing to a QSH phase are identical to those of a QWZ model reduced to one dimension \cite{osseweijer2026QWZ}, suggesting universal behavior of dimensional crossovers in topological materials.\\
For the $[110]$ ($[111]$) terminations, the corresponding phase diagrams are shown in Figs.~\ref{fig2}(g)–(j) [Figs.~\ref{fig2}(l)–(o)], with the slab unit cell for $N_z=5$ depicted in Fig.~\ref{fig2}(k) [Fig.~\ref{fig2}(p)]. The $N_z=1$ limit is excluded, as the hopping only connects sites in adjacent layers, rendering this limit a collection of isolated sites. For both $[110]$ and $[111]$ terminations, the phase diagrams display nested ellipsoidal structures similar to those of the $[100]$ case, but shifted in parameter space: they are centered at $(\Delta,m)=(0,1)$ for the $[110]$ termination and $(0,0)$ for the $[111]$ termination.

\textit{Transition to a crystalline 1D topological insulator.} We now turn to the 1D topological phase. As this phase is protected by inversion symmetry, the choice of termination becomes crucial. While multiple terminations are possible, we restrict ourselves here to $[100]$-type terminations on all surfaces. 
\begin{figure}
    \centering
    \includegraphics[width=\linewidth]{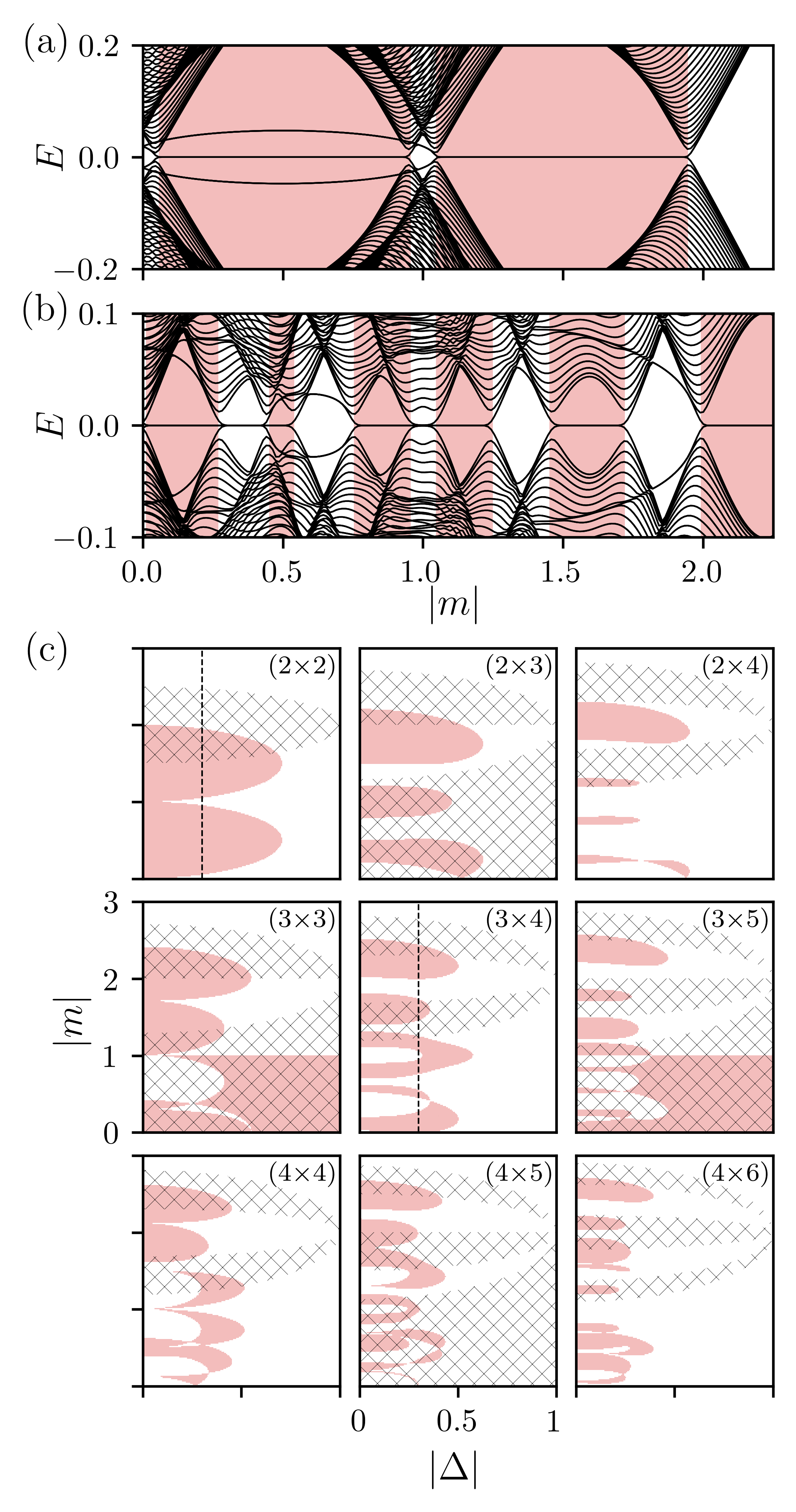}
    \caption{
(a) OBC spectrum for a $N_y \times N_z = 2 \times 2$ rod with $N_x = 100$ as a function of $m$ at $\Delta = 0.3$. The background color indicates the value of the $\mathbb{Z}_2$ invariant $\nu$, with red (white) denoting the topological (trivial) phase.
(b) OBC spectrum for a $3 \times 4$ rod with $N_x=100$ as a function of $m$ at $\Delta =0.3$.
(c) Phase diagrams as a function of $|m|$ and $|\Delta|$ for rods with different $N_y \times N_z$. Red (white) regions indicate $\nu=1$ ($\nu=0$). Hatched regions mark parameter regimes where the corresponding slab geometry at fixed $N_z$ is in the QSH phase, $\nu_\mathrm{QSH}=1$. Vertical lines indicate the parameter cuts used in (a) and (b).
}
    \label{fig3}
\end{figure}
The topological characterization can in principle be carried out using multiband Zak phases \cite{soluyanov2012smooth}. However, in the presence of inversion symmetry, a more direct and computationally efficient approach is provided by a 1D Fu-Kane-type invariant, analogous to Eq.~\eqref{eq:inv3D}. It is given by
\begin{equation}
(-1)^{\nu} = \prod_{\Lambda_i} \prod_{n}^{N_\mathrm{occ}/2} \lambda_{2n}(\Lambda_i),
\end{equation}
where the product runs over the 1D time-reversal invariant momenta $\Lambda_i = 0, \pi$, and $\lambda_n(\Lambda_i)$ are the parity eigenvalues defined through $\mathcal{I}|u_n(\Lambda_i)\rangle = \lambda_n(\Lambda_i)|u_n(\Lambda_i)\rangle$, with $\mathcal{H}_\mathrm{rod}(k)|u_n(k)\rangle = E_n(k)|u_n(k)\rangle$. Here, $\mathcal{H}_\mathrm{rod}(k)$ is the rod Hamiltonian, obtained by taking OBC in two directions. Furthermore, the $\lambda_{2n}$ makes sure that we only select one state per Kramers pair.

In Fig.~\ref{fig3}(a), we show the OBC spectrum for a long $N_y \times N_z = 2 \times 2$ rod ($N_x = 100$) as a function of $m$ at $\Delta = 0.3$. The background color indicates the value of $\nu$ for the corresponding parameters, with red denoting the topological phase and white the trivial phase. These regions are separated by bulk gap closings.
Starting from large $m$, the system enters a topological phase upon decreasing $m$ below $2$, where in-gap end states appear at $E = 0$. Around $m = 1$, the bulk gap closes and the system becomes trivial again, before reopening into a second topological phase upon further tuning of $m$. In the latter region, a pair of zero-energy end states is present. In addition, we observe finite-energy \textit{satellite states}, which are also localized at the ends of the system but disperse away from $E=0$ as a function of both $\Delta$ and $m$ and are therefore not topologically protected (see End Matter). 
In Fig.~\ref{fig3}(b), we show the corresponding OBC spectrum for a $3 \times 4$ rod. In this case, a richer sequence of topological phase transitions emerges as a function of $m$, again accompanied by satellite states. Notably, around $m \approx 0.4$, zero-energy states appear despite $\nu = 0$. This reflects the presence of two Kramers pairs at zero energy, which together constitute a topologically trivial phase due to the $\mathbb{Z}_2$ nature of the invariant.

Next, in Fig.~\ref{fig3}(c), we present phase diagrams for rods with different $N_y \times N_z$ as a function of $|m|$ and $|\Delta|$. Due to the symmetry under $N_y \leftrightarrow N_z$, only inequivalent combinations are shown. Red (white) regions indicate the topological (trivial) phase with $\nu = 1$ ($\nu = 0$). The hatched regions mark parameter regimes where the corresponding slab geometry for width $N_z$ is in the QSH phase, $\nu_\mathrm{QSH} = 1$. The parameter ranges taken in Figs.~\ref{fig3}(a) and \ref{fig3}(b) are indicated by vertical lines. The resulting phase diagrams exhibit a significantly richer structure than in the QSH case, with multiple nested topological regions as a function of both $m$ and $\Delta$. Notably, 1D topological phases can occur even in the absence of a QSH phase [e.g. a red, 1D topological, region in Fig.~\ref{fig3}(c) with no hatched region on top]. Furthermore, for rods with both $N_y$ and $N_z$ odd, a large topological region persists for $|m|<1$, analogous to the behavior of odd-layer-thick slabs in the QSH regime, see e.g. Figs.~\ref{fig2}(a),(c),(e).

\textit{Conclusions---}
We have investigated dimensional crossovers within the simplest 3D topological insulator model, focusing on slab geometries terminated along the [100], [110], and [111] crystallographic directions. For the [100] case, we further derived the phase boundaries analytically. Across all terminations, we find a strongly non-monotonic evolution of the topological character as the system is reduced in thickness, indicating that dimensional reduction does not proceed in a smooth or uniform manner\\
A central outcome of our study is the emergence of a universal structure in these crossovers: the phase diagrams obtained for the 3D-to-2D reduction closely mirror those of analogous 2D models undergoing a crossover to one dimension. This suggests that the observed behavior is not specific to a particular lattice realization, but instead reflects a broader universality class governing dimensional reduction in topological systems.\\
Upon further reduction to 1D, we identify a phase supporting Kramers-degenerate end states. We show that this 1D topological phase is protected by inversion symmetry, which plays a crucial role in stabilizing the boundary modes beyond the usual time-reversal symmetry.\\
Our findings have direct implications for a broad class of materials, such as the Bi$_2$Se$_3$ family, Pb$_{1-x}$Sn$_x$Te, HgTe quantum wells, Cd$_3$As$_2$ thin films, and elemental bismuth. While dimensional crossovers have already been explored in some of these materials, our results indicate that such transitions may extend further, potentially down to 1D limits.\\
Furthermore, our results suggest that dimensional reduction can induce topological phases even when the parent 3D material is topologically trivial. In particular, transitions from trivial 3D systems to topological 2D phases may occur. This opens the possibility that materials not classified as topological in three dimensions may nevertheless host robust topological phases in reduced dimensionality.\\
Additionally, our analysis further motivates the exploration of higher-order topological phases in this context. In particular, it would be interesting to investigate whether higher-order topological insulator phases in three dimensions can similarly undergo controlled crossovers to 1D boundary-localized states, or whether intermediate 2D QSH regimes are generically favored. Materials such as WTe$_2$ provide a promising platform for examining such intertwined dimensional and topological transitions.\\
These results establish dimensional confinement as a powerful and broadly applicable route to engineer topological phases, opening new directions for both theoretical design and experimental realization of topological states emerging from a dimensional crossover.

\textit{Acknowledgments---}
LE thanks F. van Veen for fruitful discussions. LE and CMS acknowledge the research program “Materials for the Quantum Age” (QuMat) for financial support. This program (registration number 024.005.006) is part of the Gravitation program financed by the Dutch Ministry of Education, Culture and Science (OCW).


\newpage

\section*{End Matter}
\subsection*{Evaluation of the QSH invariant}
The $\mathbb{Z}_2$ invariant is obtained from the Wilson loop spectrum evaluated over half of the slab Brillouin zone. The Wilson loop at fixed $k_1$ is defined as
\begin{equation}
    \left[\mathcal{W}_{k_1}\right]_{nm}
    =
    \prod_{j=1}^{N}
    \braket{u_n(k_1, k_2^{(j)})}{u_m(k_1, k_2^{(j+1)})},
\end{equation}
where $k_i = \mathbf{k}\cdot \mathbf{a}_i$. Here, $k_2$ is discretized along a closed loop in the Brillouin zone at fixed $k_1$, with periodic boundary condition $k_2^{(N+1)} = k_2^{(1)}$. The product is taken over the occupied bands, and $\mathcal{W}_{k_1}$ is a unitary matrix in the occupied-band subspace.

\begin{figure}[h]
    \centering
    \includegraphics[width=\linewidth]{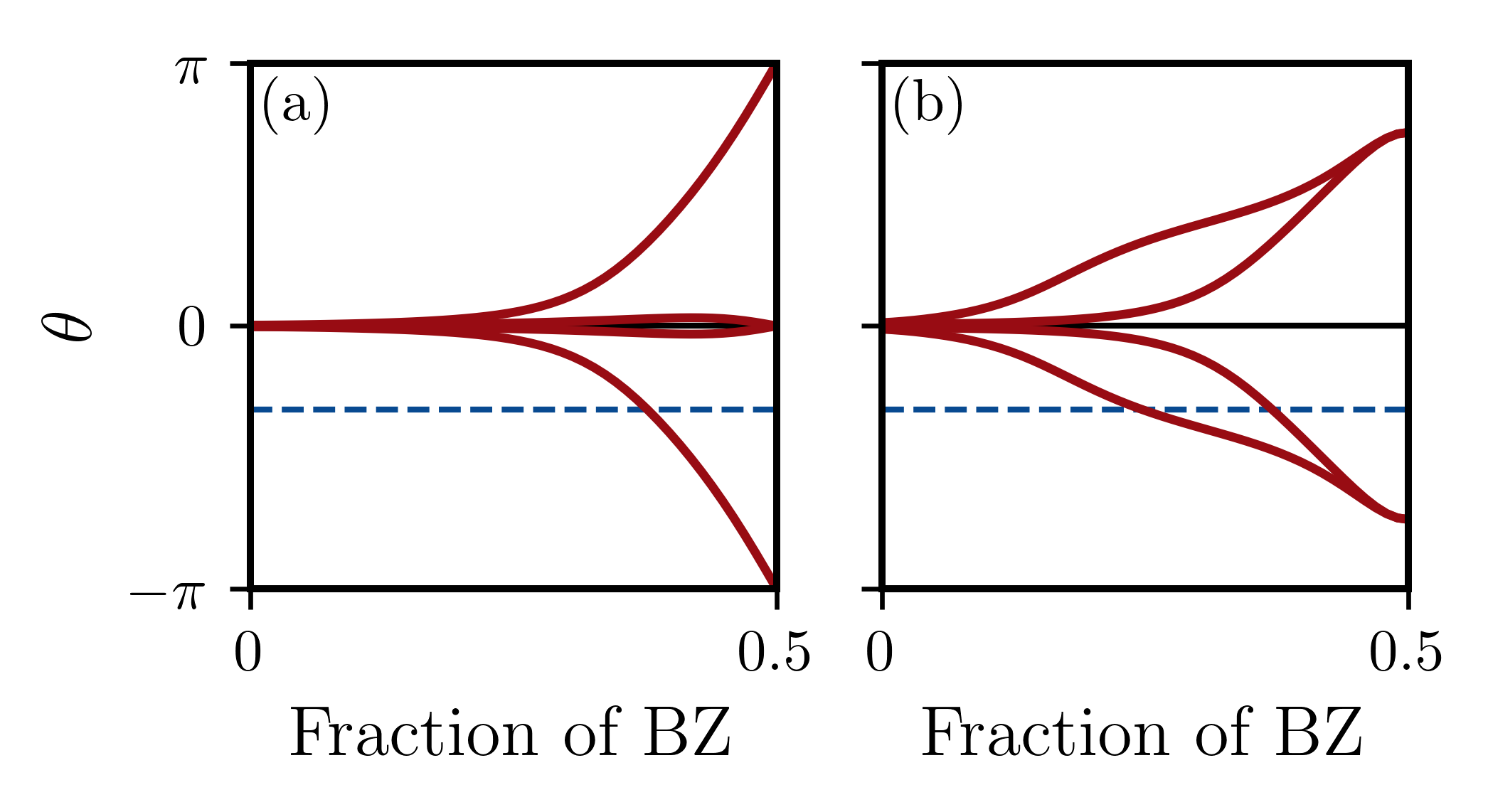}
    \caption{Wilson loop spectrum $\theta_n(k_1)$ as a function of $k_1$ for the slab Hamiltonian corresponding to Eq.~\eqref{eq:model} with $N_z=2$, with (a) $m=2$ and (b) $m=1$, both at $\Delta=1/2$. The blue dashed line denotes the reference $\theta=\mathrm{const}$.}
    \label{fig:Z2Example}
\end{figure}

The eigenvalues of $\mathcal{W}_{k_1}$ can be written as $\exp\{i\theta_n(k_1)\}$, where $\theta_n(k_1)$ are the Wilson loop phases. Evaluating $\theta_n(k_1)$ over half of the Brillouin zone, $k_1 \in [0,1/2)$, the $\mathbb{Z}_2$ invariant $\nu_{\mathrm{QSH}}$ is given by the parity of the number of crossings of the Wilson loop phases with a reference line $\theta = \mathrm{const}$ \cite{WilsonZ2}. Figure~\ref{fig:Z2Example} illustrates this procedure for both topological and trivial cases: in Fig.~\ref{fig:Z2Example}(a) [Fig.~\ref{fig:Z2Example}(b)], the Wilson loop spectrum crosses the reference line once (twice), yielding odd (even) parity and thus a topological (trivial) phase.

\subsection*{$\mathbb{Z}_2$ phase in layered Hamiltonians}
The single [100] layer of the model described in Eq.~\eqref{eq:model} is given by [written in the basis $(\ket{A\uparrow}, \ket{A\downarrow}, \ket{B\uparrow}, \ket{B\downarrow})$]
\begin{align}
    H_0(k_x,k_y) &= \left[m+t\sum_{i\in \{x,y\}} \cos (k_i) \right] \tau_z \sigma_x \notag \\ &+\Delta \left[ \sin (k_x) \tau_z \sigma_y + \sin (k_y) \tau_z \sigma_z \right].
\end{align}
The different layers are coupled together with
\begin{equation}
    H_\perp = \frac{t}{2} \tau_z \sigma_x + \frac{\Delta}{2i} \tau_x \sigma_0. 
\end{equation}
The layered Bloch Hamiltonian can then be written in the form
\begin{equation} \label{eq:blockdiag}
    H(k_x,k_y) = \begin{pmatrix}
H_0 & H_\perp \\
H_\perp^\dagger & H_0 & \ddots \\
& \ddots & \ddots & H_\perp \\
&& H_\perp^\dagger & H_0
\end{pmatrix}
\end{equation}
In the remainder of this section, we focus on $m>0$. The results for $m<0$ are readily obtained using a similar line of reasoning. In this case, the gap closings occur at the $\overline{M}$ point in the 2D Brillouin zone at $E=0$. As a result, obtaining the phase boundaries of this Hamiltonian comes down to solving $\det H(k_x=\pi, k_y=\pi)=0$. Hence, from this point onward, we will take $H_0 = (m-2t) \tau_z \sigma_x $. Equation~\eqref{eq:blockdiag} is a block-tridiagonal matrix, of which the determinant is given by \cite{molinari2008determinants}
\begin{equation}
    \det H = \prod_{n=1}^N \det \mathcal{F}_n,
\end{equation}
where $N$ is the number of layers and $\mathcal{F}_n$ is given by the recursion relation
\begin{equation}
    \mathcal{F}_n = H_0 -H_\perp^\dagger \mathcal{F}_{n-1}^{-1} H_\perp, \quad \mathcal{F}_1 = H_0.
\end{equation}
Since $\mathcal{F}_1 \propto \tau_z\sigma_x$, one finds
\begin{equation}\label{eq:alpha}
    \det \mathcal{F}_n = \alpha_n^4, \quad \alpha_n = \alpha_1-\frac{1}{4\alpha_{n-1}}(t^2-\Delta^2),
\end{equation}
where $\alpha_1 = m-2t$. Here, we recognize that $\alpha_n$ is a continued fraction, which we simplify by introducing $\alpha_n = p_n/p_{n-1}$ with $p_0=1$. Consequently, $\det H = p_N^4$ and the gap-closing condition is given by $p_N=0$. From Eq.\eqref{eq:alpha}, we obtain
\begin{equation}
    p_n = (m-2t) p_{n-1} - \frac{t^2-\Delta^2}{4}p_{n-2}.
\end{equation}
The above difference equation may be solved using the ansatz $p_n \propto \lambda^n$, yielding
\begin{equation}
    \lambda^2 - (m-2t) \lambda + (t^2-\Delta^2)/4 = 0,
\end{equation}
resulting in
\begin{equation}
    \lambda_\pm = \frac{(m-2t)\pm\sqrt{(m-2t)^2-(t^2-\Delta^2)}}{2}.
\end{equation}
We now have
\begin{equation}
    p_n = c_+ \lambda_+^n + c_-\lambda_-^n,
\end{equation}
subject to the boundary conditions ${p_0=1}$ and ${p_1 = m-2t=(\lambda_++\lambda_-)}$, which finally yields
\begin{equation}
    p_n=\frac{\lambda_+^{n+1}-\lambda_-^{n+1}}{\lambda_+-\lambda_-}.
\end{equation}
To solve this, we parametrize
\begin{equation}
    m-2t= \sqrt{t^2-\Delta^2} \cos (q), \qquad q \in \mathbb{C},
\end{equation}
such that $\lambda_\pm = \sqrt{t^2-\Delta^2} \text{exp}(\pm i q)/2$ and
\begin{figure}
    \centering
    \includegraphics[width=\linewidth]{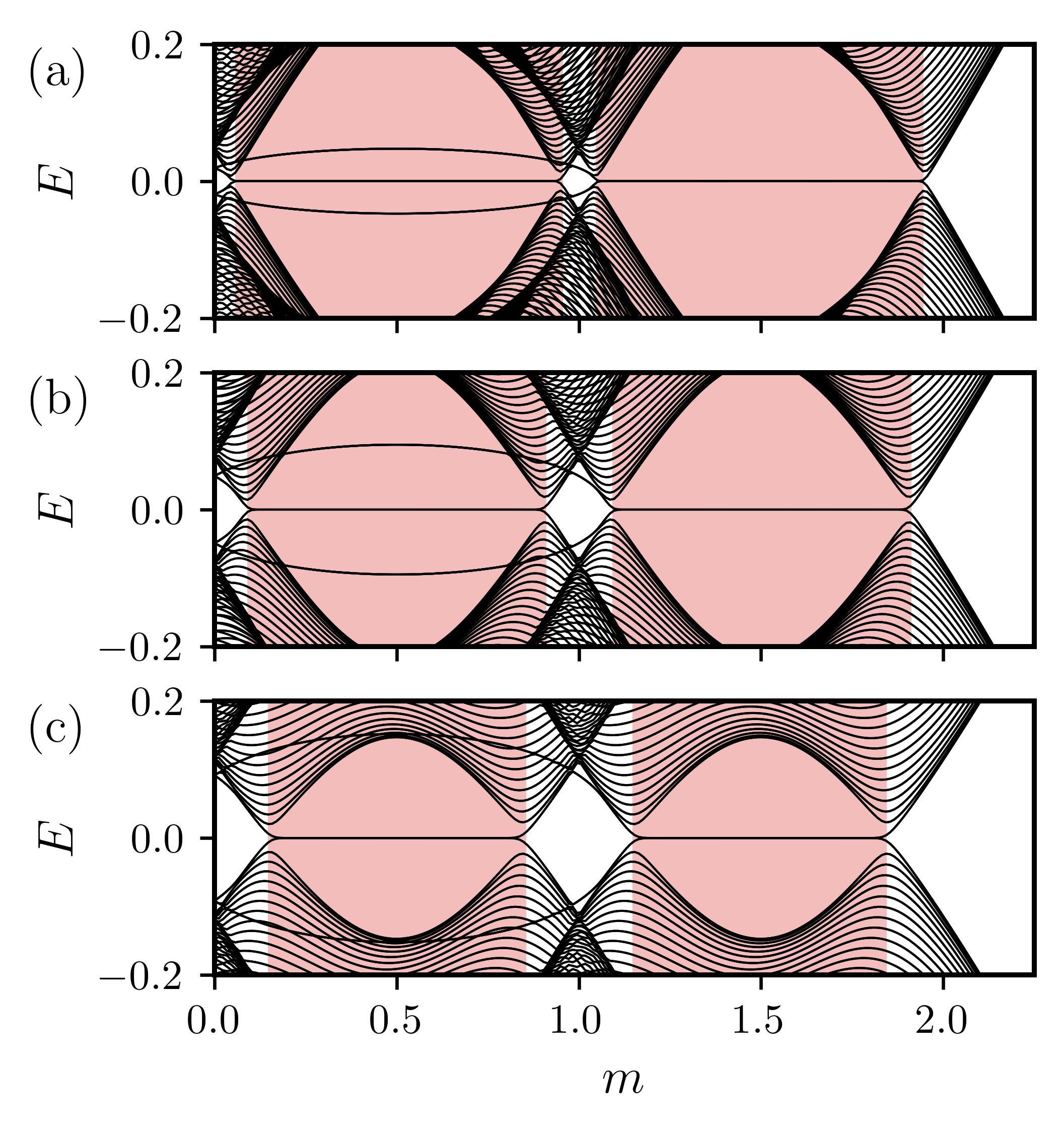}
    \caption{Spectrum of a $2\times 2\times 100$ rod Hamiltonian as a function of $m$ for (a) $\Delta = 0.3$, (b) $\Delta = 0.4$, (c) $\Delta = 0.5$.}
    \label{fig:sattelite}
\end{figure}
\begin{equation}
    p_n = \left( \frac{1}{2} \sqrt{t^2-\Delta^2} \right) ^n \frac{\sin([n+1]q)}{\sin (q)}.
\end{equation}
For a gap closing, we require $p_N=0$, which has solutions when 
\begin{equation}
    q = \frac{\pi j}{N+1} \quad j =1,\dots N.
\end{equation}
This then finally leads to the phase boundary for the [100] stacked system
\begin{equation}
    \left[\frac{m-2t}{\cos (q)}\right]^2 + \Delta^2 = t^2, 
\end{equation}
which constitutes a family of ellipses, centered at $(\Delta, m) = (0,2t)$.

\subsection*{Behavior of satellite states}
As discussed in the main text, the satellite states are deemed not topological, as they disperse with increasing $\Delta$ and $m$. In Fig.~\ref{fig:sattelite}, we show the behavior of these states for three representative choices of $\Delta$. Indeed, upon increasing $\Delta$ from $0.3$ [\ref{fig:sattelite}(a)] to $0.4$ [\ref{fig:sattelite}(b)] to $0.5$ [\ref{fig:sattelite}(c)], the states move further away from $E=0$.

\bibliography{bib}
\end{document}